# Gravitational Lensing Limits on Cold Dark Matter and Its Variants

CHRISTOPHER S. KOCHANEK

Harvard–Smithsonian Center for Astrophysics, 60 Garden St., Cambridge, MA 02138

I: kochanek@cfa.harvard.edu

## ABSTRACT

Standard $\Omega_0 = 1$ cold dark matter (CDM) needs $0.27 < \sigma_8 < 0.63$ ($2\sigma$) to fit the observed number of large separation lenses, and the constraint is nearly independent of $H_0 = 100 h^{-1}$ km s$^{-1}$ Mpc$^{-1}$. This range is strongly inconsistent with the COBE estimate of $\sigma_8 = (2.8 \pm 0.2)h$. Tilting the primordial spectrum $\propto k^n$ from $n = 1$ to $0.3 \lesssim n \lesssim 0.7$, using an effective Hubble constant of $0.15 \lesssim \Gamma = h \lesssim 0.30$, or reducing the matter density to $0.15 \lesssim \Omega_0 h \lesssim 0.3$ either with no cosmological constant ($\lambda_0 = 0$) or in a flat universe with a cosmological constant ($\Omega_0 + \lambda_0 = 1$) can bring the lensing estimate of $\sigma_8$ into agreement with the COBE estimates. The models and values for $\sigma_8$ consistent with both lensing and COBE match the estimates from the local number density of clusters and correlation functions. The conclusions are insensitive to systematic errors except for the assumption that cluster core radii are singular. If clusters with $\rho \propto (r^2 + s^2)^{-1}$ have core radii exceeding $s = 15 h^{-1} \sigma_3^2$ kpc for a cluster with velocity dispersion $\sigma = 10^3 \sigma_3$ km s$^{-1}$ then the estimates are invalid. There is, however, a fine tuning problem in making the cluster core radii large enough to invalidate the estimates of $\sigma_8$ while producing several lenses that do not have central or "odd images." The estimated completeness of the current samples of lenses larger than $5\farcs0$ is 20%, because neither quasar surveys nor lens surveys are optimized to find this class of lenses.

*Subject headings:* gravitational lensing – cosmology: observations – cosmology: theory – dark matter – large-scale structure of universe

## 1  INTRODUCTION

Narayan & White (1988) pointed out that the standard cold dark matter (CDM) model predicts many more gravitational lenses with separations above $5\farcs0$ than we find. Their predictions were based on the Press-Schechter (1974, PS hereafter) model, but more recent numerical models by Cen et al. (1994) and Wambsganss et al. (1994) confirm the fundamental result. A singular isothermal sphere with velocity dispersion $\sigma = 10^3 \sigma_3$ km s$^{-1}$ produces lenses with an average image separation of $\Delta\theta = 28\rlap{.}''8\sigma_3^2$, so the large separation lenses explore the number and evolution of clusters and groups. They are, however, a qualitatively different test of cosmogonic models than the local density of clusters (Peebles et al. 1989, Frenk et al. 1990, Bahcall & Cen 1992, 1993) or correlation functions (Maddox et al. 1990, Picard 1991, Vogeley et al. 1992, Loveday et al. 1992). Any massive, collapsed, virialized halo will produce lenses, so the test is independent of the luminosity of the lenses and unaffected by problems with detecting and counting complete cluster samples locally. The probability of lensing peaks at intermediate redshifts, and it goes to zero at low redshift, so it is a test of the number density of groups and clusters at $z \sim 0.3\text{-}0.5$ rather than at $z = 0$. This allows lensing to distinguish between scenarios that produce the same number of clusters today using different formation histories.

If we can understand the selection function for large separation lenses in heterogeneous quasar catalogs, compute the magnification bias of the sample, and decide which of the large separation quasar pairs to call lenses, then we have an important new cosmological probe. Only qualitative comparisons can be made without including magnification bias and selection effects, because they can change the number of lenses found in any observational sample by an order of magnitude. Unfortunately, Narayan & White (1988), Cen et al. (1994), and Wambsganss et al. (1994) only made qualitative comparisons between the models and the observations. Narayan & White (1988) and Wambsganss et al. (1994) did not include selection effects and magnification bias, and Cen et al. (1994) used a crude model based on lens surveys. Where Cen et al. (1994) and Wambsganss et al. (1994) pursued numerical calculations of cross sections, we will focus on selection effects, magnification bias, and quantitative estimates of the number of lenses in various cosmological scenarios. The disadvantage of our approach is that, like Narayan & White (1988), we rely on the PS model to estimate the number and distribution of lenses. We can, however, see if the PS formalism is an accurate method for estimating lens probabilities by comparing to the Wambsganss et al. (1994) numerical results when possible and by examining the effects of the systematic uncertainties on the conclusions. The advantage of the PS model is that we can rapidly survey a large number of cosmological scenarios to examine the sensitivity of the method to its parameters and to see which models are constrained by gravitational lensing,

We know of two confirmed lenses with separations larger than $3\rlap{.}''0$ (Q0957+561 and Q2016+112), and another four candidates (Q1120+019=UM 425, Q1429−008, Q1635+267, and Q2345+007). Two additional pairs, PKS 1145−071AB (Djorgovski et al. 1987) and Q1343+266AB (Crampton et al. 1988), are rejected as lens candidates even though the redshift difference in both pairs is less than $\Delta z \leq 0.001$. PKS 1145−071 is rejected because one quasar is radio loud and the other is radio quiet (>500:1 flux ratio), and Q1343+266AB is rejected because of gross differences in the spectral lines of the quasars. The properties of these eight objects are summarized in Table 1. A key distinction between the pairs in Table 1 is whether they were found as part of the original survey that found the quasar, or whether they were found in a lens survey examining known quasars to see if they are lensed. Four of the eight objects in Table 1 were found in the original quasar survey (Q0957+561, Q1343+266, Q1635+267, and Q2345+007) and four were found as part of a search for lensed images (Q1120+019, PKS 1145−071, Q1429−008, and Q2016+112). Q1120+019 and PKS

1145−071 were found in a survey by Djorgovski & Meylan (1989), Q1429−008 in a survey by Webster et al. (1988), and Q2016+112 in the MG survey (Burke et al. 1992). In the first two cases the survey lists and selection functions are unpublished, so we cannot build a theoretical model. For Q2016+112 we do not have the necessary information on the redshift and radio flux distributions of the MG sources to make a theoretical model. For comparison, all the galaxy scale lenses but one (PG1115+080, Weymann et al. 1980) were found in lens surveys.

The key to drawing quantitative conclusions is §2, where we develop a selection effects model for finding lenses in a heterogeneous quasar like the Hewitt-Burbidge (1993, HB93 hereafter) catalog. More importantly, we show that it is a valid selection effects model for lensed quasars from the statistical properties of the unlensed objects in the catalog. In §3 we summarize the theory of gravitational lens statistics and discuss the effects of the selection model on the probability that a lens is detectable. In §4 we review and expand the PS lensing model developed by Narayan & White (1988). In §5 we examine the standard CDM model, and some of the variants suggested to correct the problems in COBE normalized CDM. In §6 we consider sources of systematic error in the calculation and how they limit the cosmological constraints, and in §7 we review the results and discuss the requirements for better wide separation lens surveys.

## 2 A Selection Effects Model for Heterogeneous Quasar Catalogs

We examine the statistics of large separation lenses in the HB93 catalog, and Table 1 summarizes all the known lensed pairs, candidate pairs, and associated quasars with separations larger than $3''\!.0$ in the HB93 catalog or commonly appearing in lists of gravitational lenses. The known galaxy scale lenses with separations smaller than $3''\!.0$ (see review by Surdej & Soucail 1994) and the eight objects in Table 1 are the full sample of objects in the HB93 catalog with redshifts above 1.0, redshift differences smaller than 0.01, and separations smaller than $1'$.[1].

Wide separation lenses are always resolved, so given one image of the lens with magnitude $m_1$ brighter than the magnitude limit $m_l$, the selection function requires that any second lensed image with magnitude $m_2$ must also be brighter than the survey magnitude limit. The magnitude limit varies from survey to survey, but we would like to have a plausible model for the average dynamic range between a quasar in a heterogeneous catalog and its magnitude limit. We can do this based on two plausible assumptions: (1) all surveys have magnitude limits, and (2) surveys for fainter quasars do not substantially overlap surveys for brighter

---

[1] Be warned, however, that a search of the HB93 catalog with our selection criteria will find seventeen additional pairs. Six of these "pairs" (two pairs near NGC 3384 (Q1045+128, Arp et al. 1979), one pair near NGC 2683 (Q0849+336, Arp 1983), and three near Q1549+486 (Arp & Surdej 1982)) have separations much larger than one arc-minute. The discovery papers did not give separate positions for the quasars in the field, so all of the quasars in each field were given a common position. Eleven other quasars (Q0057−352, Q0059−411, Q0252+016, Q0255−015, Q0851+197=LB8863, Q0953+549, Q1128+105, Q1151+068, Q1208+142, Q1209+107, and Q1213+155) appear to be double entries of the same quasar.

Table 1: Quasar Lenses and Pairs With Separations Larger Than 3″.0

| Name | $z_s$ | $\Delta\theta$ | $m_A$ | $m_B$ | Band | Lens? | Why | Who |
|---|---|---|---|---|---|---|---|---|
| Q0957+561 | 1.41 | 6″.1 | 17.5 | 17.7 | B | Yes | Q | Walsh et al. 1979 |
| Q2016+112 | 3.27 | 3″.6 | 22.9 | 23.2 | i | Yes | L | Lawrence et al. 1984 |
| Q1120+019 | 1.46 | 6″.5 | 16.2 | 20.8 | B | ? | L | Meylan & Djorgovski 1989 |
| Q1429−008 | 2.08 | 5″.1 | 17.7 | 20.8 | R | ? | L | Hewett et al. 1989 |
| Q1635+267 | 1.96 | 3″.8 | 19.2 | 20.8 | B | ? | Q | Djorgovski & Spinrad 1984 |
| Q2345+007 | 2.15 | 7″.3 | 19.5 | 21.0 | B | ? | Q | Weedman et al. 1982 |
| PKS1145−071 | 1.35 | 4″.2 | 18.0 | 18.8 | B | No | L | Djorgovski et al. 1987 |
| Q1343+264 | 2.03 | 9″.5 | 20.8 | 20.9 | B | No | Q | Crampton et al. 1988 |

Notes: The entries in the Lens? column are Y if the object is generally believed to be a lens, N if it is generally believed not to be a lens, and ? if its status is uncertain. The entries in the Why column are Q if the object was found as part of a quasar survey, and L if the object was found as part of a lens survey.

quasars. The first assumption is trivial, but the second requires some justification. The surface density of quasars is a steep function of magnitude (the surface densities of quasars brighter than 15, 19, and 21 B mags are $1.7 \times 10^{-3}$, 4.3, and 33 per square degree for the redshift range $0 < z < 2.2$ (Hartwick & Schade 1990), so to find equal numbers of quasars, bright quasar surveys cover large areas and faint quasar surveys cover small areas. Since the total area surveyed for faint quasars is much smaller than that surveyed for bright quasars, the typical bright quasar is not part of a faint quasar survey. Thus the magnitude limit for finding companions to any quasar is determined by the magnitude limit of the survey that found the quasar.

We consider a model survey for quasars with a limiting magnitude $m_l$. We assume that the survey finds all quasars brighter than the limiting magnitude in a region much larger than the largest interesting lens separation. The model neglects photometric errors and Eddington bias (see Hartwick & Schade 1990). We model the quasar apparent magnitude number counts by a broken power law

$$\frac{dN}{dm} = N_0 \begin{cases} 10^{\alpha(m-m_0)} & m < m_0 \\ 10^{\beta(m-m_0)} & m > m_0 \end{cases} \qquad (1)$$

where $\alpha \simeq 1.12$, $\beta \simeq 0.18$, and $m_0 \simeq 19.1$ B mags (Hartwick & Schade 1990, Boyle et al. 1990, Wallington & Narayan 1993). The apparent magnitude of the break $m_0$ is nearly constant for the redshift range $1 < z < 3.5$. In this paper we are uninterested in the absolute normalization of the number of quasars $N_0$. The probability distribution for the magnitude difference $\Delta m = m_l - m$ between a survey quasar with magnitude $m$ and the magnitude limit is

$$\frac{dP}{d\Delta m} = \left[\int_{-\infty}^{m_l} \frac{dN}{dm} dm\right]^{-1} \frac{dN}{dm}(m_l - \Delta m) \qquad \text{for} \qquad \Delta m > 0. \qquad (2)$$

For surveys with magnitude limits brighter than the break magnitude, $m_l < m_0$, the differ-ential and integral probability distributions take the simple forms

$$\frac{dP}{d\Delta m} = [\alpha \ln 10] \, 10^{-\alpha \Delta m} \quad \text{and} \quad P(<\Delta m) = 1 - 10^{-\alpha \Delta m}. \tag{3}$$

In this limit, the mean dynamic range is $\langle \Delta m \rangle = (\alpha \ln 10)^{-1} = 0.34$ mag, the median is $\Delta m_{1/2} = \alpha^{-1} \log 2 = 0.24$ mag, and 90% of the quasars have $\Delta m < \alpha^{-1} = 0.79$ mag. The dynamic range between a quasar and the magnitude limit in any magnitude limited bright quasar survey is very small. When the magnitude limit becomes fainter than the break magnitude $m_0$, the average dynamic range increases. In the limit that $m_l \gg m_0$ the mean dynamic range is $\langle \Delta m \rangle = (\beta \ln 10)^{-1} = 1.55$ mag, the median is $\Delta m_{1/2} = \beta^{-1} \log 2 = 1.67$ mag, and 90% of the quasars have $\Delta m < \beta^{-1} = 5.5$ mag. For comparison, surveys for lensed quasars (eg. Maoz et al. 1993, Surdej et al. 1993) have typical dynamic ranges of 4 to 5 magnitudes.

We test this model by examining the magnitude differences between pairs of quasars in the HB93 catalog. We took the ∼ 5000 quasars with measured V magnitudes and $1 < z < 4$ in the HB93 catalog, and found the nearest neighbor for each quasar excluding known lenses and the pairs in Table 1. Figure 1 shows the mean and the dispersion of the magnitude differences of the pairs as a function of the V magnitude of the brighter quasar. The pairs are collected in bins one magnitude wide that are subdivided into a maximum of three smaller bins as the number of quasars in the bin increases. Figure 1 also shows the mean and dispersion predicted for $\Delta m$ by the selection function model assuming each quasar is at the median magnitude for its magnitude limit, and an average $B - V$ color of 0.2 mag. Thus, a 18 V mag quasar is assumed to come from a survey with a magnitude limit of $m_l = 18 + \alpha^{-1} \log 2 = 18.24$ V mag. The agreement between the data and the model for both the average magnitude difference and the spread in the magnitude difference is remarkably good.

Figure 1 also shows the locations of the eight quasar pairs in Table 1, labeled by the type of survey that found the pair. As expected, the large magnitude difference pairs (Q1120+019 and Q1429−008) were found as part of a lens survey. The four pairs found in the quasar surveys roughly follow the expected selection function. Although PKS 1145−071 was found in a lens survey, it was selected because its image was visibly elongated in the original quasar finding chart (Djorgovski et al. 1987). This explains why it lies in the range detectable by the original quasar survey. The distribution of lensed pairs will not have the same statistical properties as the distribution of unlensed pairs because of magnification bias.

## 3  The Effect of the Selection Function on Finding Lenses

We examine the effect of the selection model on the expected number of lenses using the simple singular isothermal sphere (SIS) model for the lenses. A SIS lens with velocity dispersion $\sigma$ produces two images with angular separation $8\pi(\sigma/c)^2 D_{LS}/D_{OS}$ where $D_{LS}$ and $D_{OS}$ are the proper motion distances between the lens and the source and the observer and the source respectively. The integral probability distribution for the two images having a total magnification larger than $M$ is $P(>M) = 4/M^2$ with $M \geq 2$, and the flux ratio between the two images is $f = (M - 2)/(M + 2)$ with $0 \leq f \leq 1$ (Gott & Gunn 1974, Turner, Ostriker, & Gott 1984). The integral probability for finding two images with a flux ratio larger than $f$ (closer to unity) is

$$P(>f) = \left[\frac{1-f}{1+f}\right]^2. \tag{4}$$

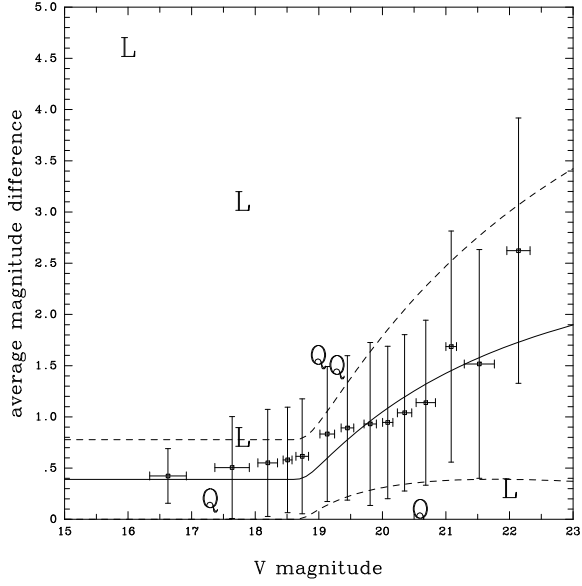

Fig. 1.—Average (points) and rms scatter (vertical error bars) of nearest neighbors with redshifts $1 < z < 4$ in the HB93 catalog as a function of magnitude The horizontal error bars show the width of the magnitude bins. The lines show the predicted average (solid) and rms (dashed) width of the distribution predicted by the model selection function. The quasar pairs in Table 1 are marked by a Q if the pair was found in a quasar survey, and an L if the pair was found in a lens survey.

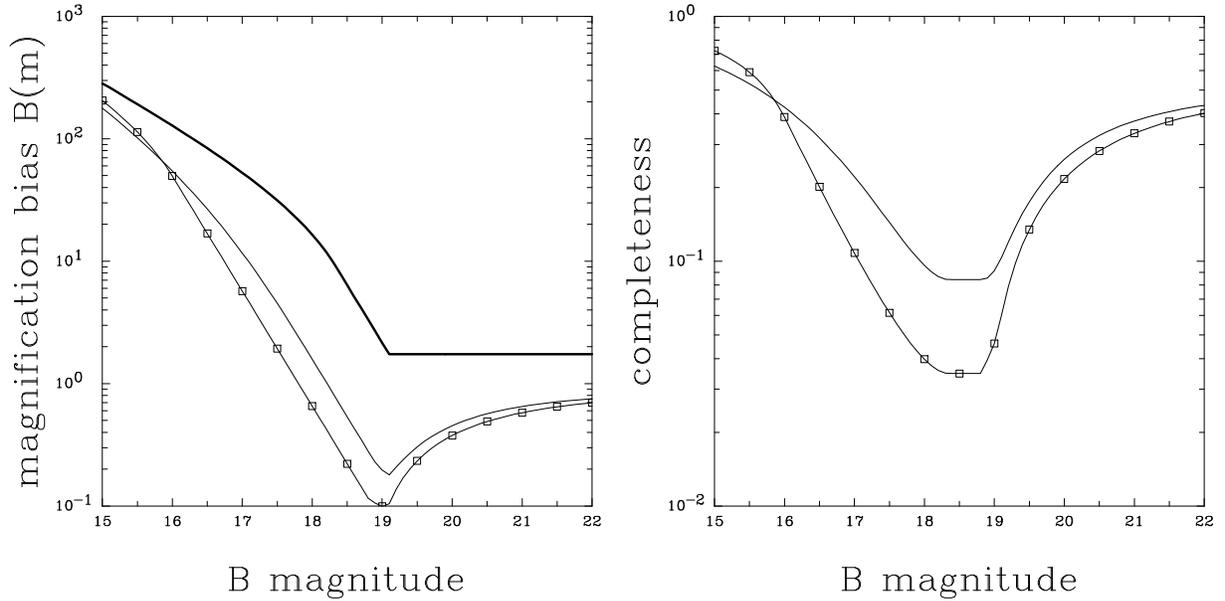

Fig. 2.—Magnification bias as a function of B magnitude. The left panel shows the bias factor if we find all lenses (heavy solid), all lenses with magnitude differences smaller than the median for the selection function (light solid/points), and the average of the bias factor over the distribution of magnitude differences (light solid). The right panel shows the completeness, or the ratio of the magnification bias with a selection function to the magnification bias when we find all lenses.

For bright quasars the median dynamic range in our model for the selection function is $\Delta m_{1/2} = 0.24$, so we only find lenses with $f > f_l = -2.5 \log \Delta m_{1/2} = 0.80$ and $P(>f_l) = 0.012$. If we average $P(>f)$ over the distribution of dynamic range (eqn. 2), we find an average of $P(>f_l) = 0.043$ of the lensed bright quasars. For comparison, Wambsganss et al. (1994) assumed a uniform dynamic range of 1.5 magnitudes, for which $P(>f_l) = 0.36$. In the SIS model, this assumption overestimates the expected number of bright lenses by a factor of 7. If we relied on the lens cross section to determine the number of lenses, lensing would be useless as a cosmological test.

Fortunately, predictions of the number of lenses found in real surveys must include the effects of magnification bias to correct for the difference between the number of quasars at the magnitude of the unlensed source and the number of quasars at the magnitude of the lensed source (Gott & Gunn 1974, Turner 1980). The bias factor is

$$B(m) = \left[\frac{dN}{dm}(m)\right]^{-1} \int_{M_{lim}}^{\infty} dM \frac{8}{M^3} \frac{dN}{dm}(m + 2.5 \log M) \tag{5}$$

where $m$ is the magnitude of the quasar, $M$ is the total magnification, and $M_{lim}$ is the magnification at which the images have the minimum detectable flux ratio (Fukugita & Turner 1991, Kochanek 1991). The completeness, or fraction of the lenses we can detect at a given magnitude, is the ratio of the bias factor with $M_{lim} = 2(1 + f_l)/(1 - f_l)$ and the bias factor with $M_{lim} = 2$. Figure 2 shows the magnification bias assuming we find all lenses $B_T(m)$, all lenses with flux ratios larger than the median dynamic range predicted by the selection function, $B_{1/2}(m)$, and the average of the magnification bias over the probability distribution for the dynamic range (eqn. 2), $\langle B \rangle(m)$. The magnification bias compensates for the low optical depth at most magnitudes, although the Cen et al. (1994) model significantly overestimates the amount of bias. The standard quasar number counts used by Cen et al. (1994) based on Fukugita & Turner (1991) also have a significantly shallower slope for the number counts of the bright quasars ($\alpha = 0.86$, $\beta = 0.28$) than more recent determinations ($\alpha = 1.12$, $\beta = 0.18$; see Boyle et al. 1990, Wallington & Narayan 1993).

Given the selection function for finding lensed quasars and the magnitude limit averaged magnification bias factor $\langle B \rangle(m)$, the probability that a quasar with redshift $z$ and B magnitude $m$ is a lens with an image separation larger than $\Delta\theta$ is

$$p(m, z, > \Delta\theta) = 16\pi^3 \langle B \rangle(m) \int_0^{D_{OS}} \frac{D_{OL}^2 dD_{OL}}{(1 + \Omega_K D_{OL}^2 r_H^{-2})^{1/2}} \left(\frac{D_{LS}}{D_{OS}}\right)^2 \int_{\sigma_{min}}^{\infty} d\sigma \left(\frac{\sigma}{c}\right)^4 \frac{dn}{d\sigma}(\sigma, z) \tag{6}$$

where $dn/d\sigma(\sigma, z)$ is the comoving number density of lenses with velocity dispersion $\sigma$ at redshift $z$, $\sigma_{min} = c(\Delta\theta D_{OS}/8\pi D_{LS})^{1/2}$ is the smallest velocity dispersion that can produce an image separation $\Delta\theta$ at redshift $z$, and $c$ is the speed of light. The distances $D_{OS}$, $D_{OL}$, and $D_{LS}$ are the proper motion distance to the source, to the lens, and between the lens and the source. The Hubble radius is $r_H = c/H_0$, $H_0 = 100h$ km s$^{-1}$ Mpc$^{-1}$, and $\Omega_K = 1 - \Omega_0 - \lambda_0$ is the "curvature density" for a cosmological model with matter density $\Omega_0$ and cosmological constant $\lambda_0$ (Carroll et al. 1990, Kochanek 1993a). The distance $D_{LS}$ can be determined from $D_{OL}$ and $D_{OS}$ by the relation $D_{LS} = D_{OS}(1 + \Omega_K D_{OL}^2 r_H^{-2})^{1/2} - D_{OL}(1 + \Omega_K D_{OS}^2 r_H^{-2})^{1/2}$. From equation (6) we can also compute the probability that a lens has separation $\Delta\theta$ to be

$$\frac{dP}{d\Delta\theta} = \frac{\pi}{8}\Delta\theta \langle B\rangle(m) \int_0^{D_{OS}} \frac{D_{OL}^2 dD_{OL}}{(1 + \Omega_K D_{OL}^2 r_H^{-2})^{1/2}} \sigma \frac{dn}{d\sigma} \qquad (7)$$

where $\sigma = c(\Delta\theta D_{OS}/8\pi D_{LS})^{1/2}$. The integrand of $dP/d\Delta\theta$ (converted to a differential with respect to lens redshift) is the lens redshift probability distribution for lenses with image separation $\Delta\theta$.

Given the number density of potentials $dn/d\sigma$ we compute the probability $p_i$ that the $i^{th}$ quasar with B magnitude $m_i$ and redshift $z_i$ in the HB93 catalog is lensed and detectable given our model selection function. We use the maximum likelihood formalism of Kochanek (1993b) to estimate the likelihood that different models fit the lens data. For a sample of $N_U$ unlensed quasars and $N_L$ lensed quasars the likelihood $L$ of the observations is

$$\ln L = -\sum_{k=1}^{N_U} p_k + \sum_{i=1}^{N_L} \ln p_i \qquad (8)$$

where we use the expansion $\ln(1 - p_k) = -p_k$ because $p_k \ll 1$. If $L_{max}$ is the maximum value of the likelihood for some range of model parameters, then the function $-2\ln(L/L_{max})$ is asymptotically distributed like the $\chi^2$ distribution (Lupton 1993). The 68% ($1\sigma$), 90%, 95.4% ($2\sigma$) and 99% confidence levels on parameters in one (two) dimensions are where the likelihood is 60.7% (31.7%), 25.8% (9.98%), 13.5% (4.57%), and 3.63% (1.00%) of the peak likelihood.

We present two standard calculations. The first is the likelihood of finding lenses larger than $\Delta\theta = 5\rlap{.}''0$, and the second is the likelihood of finding lenses larger than $3\rlap{.}''0$ including the likelihood of finding the lenses with their observed separations (eqn. 7 instead of eqn. 6). By examining these two different statistical estimates we check whether the conclusions are sensitive to the lower angular cutoff ($5\rlap{.}''0$ versus $3\rlap{.}''0$) and whether the results are strongly sensitive to the observed separations. We know that $N_L$ is at least equal to one, because there is one certain lens in the sample, Q0957+561. There might be as many as three lenses larger than $5\rlap{.}''0$ if we include Q2345+007 and Q1343+007. We generally present results for $N_L = 1$ and $N_L = 3$ to show the sensitivity of the conclusions to the ambiguities in the numbers of lenses. When we use the $3\rlap{.}''0$ angular limit we drop the Q1343+007 pair and replace it with the lens candidate Q1635+267.

## 4 Estimates of the Number Density of Potentials

Following the approach of Narayan & White (1988) we use the PS model to estimate the comoving number density of clusters $dn/d\sigma$ as a function of redshift $z$ and velocity dispersion $\sigma$. Let $\delta_c(z)$ be the critical overdensity $\delta\rho/\langle\rho\rangle$ that can collapse before redshift $z$, and assume that the fluctuations at a comoving scale of $r_0$ are Gaussian with an rms linear overdensity of $\Delta(r_0)$. The fraction of the universe in such high density regions is $F(r_0, z) = (2\pi\Delta(r_0))^{-1/2} \int_{\delta_c(z)}^{\infty} \exp(-u^2/2\Delta(r_0)^2) du$. PS suggested that the number of clumps with initial comoving radii in the range $r_0$ to $r_0 + dr_0$ can be approximated by $f(r_0, z) dr_0 = -2(\partial F/\partial r_0) dr_0$, where the 2 is inserted to ensure that $\int_0^{\infty} f dr_0 = 1$. Perturbations exceeding the critical overdensity collapse and virialize. Bond et al. (1991) give a rigorous derivation of the PS *ansatz*. After multiplying $f(r_0, z)$ by the appropriate Jacobian and normalizing it, we find that the comoving number density of halos at redshift $z$ is

$$\frac{dn}{d\sigma}(\sigma, z) = \frac{-3\delta_c(z)}{(2\pi)^{3/2} r_0^3 \sigma \Delta} \frac{d\ln\Delta}{d\ln r_0} \exp\left[-\frac{\delta_c^2(z)}{2\Delta^2}\right] \quad (9)$$

if the velocity dispersion of the collapsed object is $\sigma \propto r_0$. If the power spectrum of the fluctuations is $|\delta_k|^2$ then the variance of the fluctuations on scale $r_0$ is given by the convolution

$$\Delta^2(r_0) = (2\pi)^{-3} \int_0^{\infty} 4\pi k^2 dk |\delta_k|^2 W^2(kr_0) \quad \text{where} \quad W(x) = 3\left[\frac{\sin x - x\cos x}{x^3}\right] \quad (10)$$

is the Fourier transform of the top-hat window function. We normalize the power spectrum by the rms fluctuation $\sigma_8 = \Delta(r_0 = 8h^{-1}\text{Mpc})$ on the scale of $r_0 = 8h^{-1}$ Mpc, or by the "bias" factor $b = 1/\sigma_8$. For any cosmological model and power spectrum we can compute the expected number of lenses given the function $\delta_c(z)$ for the smallest perturbation that can collapse at redshift $z$ and the relationship between $\sigma$ and $r_0$.

Bartelmann et al. (1993) and Lacey & Cole (1993) show that the redshift $z$ at which an initial perturbation with amplitude $\delta_i$ at redshift $z_i \gg \Omega_0^{-1}$ collapses in a $\lambda_0 = 0$ cosmological model is

$$\delta_i(z) = \frac{3}{5} \frac{1}{\Omega_0(1+z_i)} \left[\left(\frac{\pi\Omega_0}{F_1(z)}\right)^{2/3} + 1 - \Omega_0\right] \quad (11)$$

where the age of the universe at redshift $z$ is

$$T(z) = H_0^{-1} F_1(z) = H_0^{-1} \left[\frac{(1+\Omega_0 z)^{1/2}}{(1-\Omega_0)(1+z)} - \frac{\Omega_0}{2(1-\Omega_0)^{3/2}} \cosh^{-1}\left\{\frac{2(1-\Omega_0)}{\Omega_0(1+z)} + 1\right\}\right]. \quad (12)$$

We use a power spectrum that is normalized by the fluctuations today, so we must use linear theory to convert from $\delta_i$ to $\delta_c$. The growing mode changes its amplitude with redshift by the function (Peebles 1980)

$$D_1[x] = 1 + \frac{3}{x} + \frac{3(1+x)^{1/2}}{x^{3/2}} \ln[(1+x)^{1/2} - x^{1/2}] \quad \text{where} \quad x = \frac{\Omega_0^{-1} - 1}{1+z}, \quad (13)$$

so

$$\delta_c(z) = \frac{3}{2} \frac{D_1[\Omega_0^{-1} - 1]}{1 - \Omega_0} \left[\left(\frac{\pi\Omega_0}{F_1(z)}\right)^{2/3} + 1 - \Omega_0\right]. \quad (14)$$

In the limit that $\Omega_0 \to 1$ the function $D_1[\Omega_0^{-1} - 1]/(1 - \Omega_0) \to (2/5)$ and $F_1(z) \to (2/3)(1+z)^{-3/2}$, and we recover the standard result that $\delta_c(z) = (3/5)(3\pi/2)^{2/3}(1+z)$.

The mass of the collapsing perturbation is $M = 4\pi\rho_0 r_0^3/3$ where $\rho_0 = 3H_0^2\Omega_0/8\pi G$ is the average mass density. The object collapses and virializes to form an isothermal sphere of velocity dispersion $\sigma$, radius $R_f$, and a mass inside $R_f$ of $M = 2\sigma^2 R_f/G$. Equating the two masses we find that

$$\sigma = H_0 r_0 \Omega_0^{1/2} \left[\frac{r_0}{4R_f}\right]^{1/2} \tag{15}$$

The virial theorem for the collapsing perturbation shows that the final virialized radius is $R_f = R_{max}/2$ (Lahav et al. 1991), where $R_{max} = r_0(1+z_i)^{-1}(5\delta_i/3 - \epsilon_i)^{-1}$. Combining the expression for $R_{max}$ with equation (15), we find that

$$\sigma = H_0 r_0 \Omega_0^{1/3} 2^{-1/2} \pi^{1/3} F_1(z)^{-1/3}. \tag{16}$$

In the limit that $\Omega_0 = 1$ we find that standard result that $\sigma = 2^{-1/2}(3\pi/2)^{1/3} H_0 r_0 (1+z)^{1/2}$.

For flat universes with a positive cosmological constant ($\Omega_0 + \lambda_0 = 1$) none of the integrals needed for the PS model can be done analytically, but following Richstone, Loeb, & Turner (1992) we numerically solve the implicit equations for $\delta_c$. The age of the universe $T = H_0^{-1} F_1(z)$ is given by

$$F_1(z) = \int_0^{(1+z)^{-1}} u^{1/2} du \left[u^3 \lambda_0 + \Omega_0\right]^{-1/2} \tag{17}$$

and the amplitude of the growing mode is proportional to

$$D_1(z) = \left[\lambda_0 + (1+z)^3 \Omega_0\right]^{1/3} \int_0^{(1+z)^{-1}} u^{3/2} du \left[u^3 \lambda_0 + \Omega_0\right]^{-3/2}. \tag{18}$$

The collapse time $\tau_c(\delta_i)$ for a perturbation with fractional overdensity $\delta_i$ at redshift $z_i \gg 1$ is

$$H_0 \tau_c(\delta_i) = 4\lambda_0^{-1/2} \int_0^{\pi/2} \sin^2\theta \left[u_2/u_1 - \sin^2\theta\right]^{-1/2} \left[\sin^2\theta - u_3/u_1\right]^{-1/2} \tag{19}$$

where $u_3 < 0 < u_1 < u_2$ are the roots of the cubic polynomial in $u = r/r_0$

$$0 = 1 - \epsilon_i + \delta_i - 5\delta_i u/3 + \epsilon_i u^3 \tag{20}$$

and $\epsilon_i = \lambda_0 \left[\lambda_0 + \Omega_i(1+z_i)^3\right] \ll 1$ is the cosmological constant or the deviation of the matter density from unity ($\Omega_i = 1 - \epsilon_i$) at redshift $z_i$. The root $u_1 = R_{max}/r_0$ is the ratio of the maximum radius $R_{max}$ of a perturbation to its initial radius $r_0$, and the perturbation must be larger than $\delta_i \gtrsim (9/5)(\epsilon_i/4)^{1/3}$ to collapse in finite time. The function $\delta_c(z)$ is constructed by solving the implicit equation $F_1(z) = H_0 \tau_c(\delta_i)$ for $\delta_i(z)$ and using the growth rate of the growing mode to convert from $\delta_i$ to $\delta_c(z) = [D_1(0)/D_1(z_i)]\delta_i(z)$.

The velocity dispersion of the collapsed object in the models with a cosmological constant must correct for the change in the virial radius from the energy associated with the cosmological constant (Lahav et al. 1991). If $\eta = \lambda/4\pi G\rho_{ta}$ is the ratio of the cosmological constant to the average density when the perturbation turns around and collapses, then the ratio of the virial radius $R_f$ to the turn around radius $R_{max}$ satisfies the cubic equation $2\eta(R_f/R_{max})^3 - (2+\eta)(R_f/R_{max}) + 1 = 0$. When $\eta = 0$ we find the standard result $R_f/R_{max} = 1/2$, but when $\lambda_0 > 1$ the lack of the repulsive force from a positive cosmological constant in the collapsed halo leads to a smaller virialized object. The solution for the virial radius is well approximated by $R_f/R_{max} = 0.5 - 0.138\eta + 0.0034\eta^2$ for $\eta > 0$. (For $\eta > 0$ this is more accurate than the approximation in Lahav et al. (1991).)

## 5  Lensing in Different Cosmological Scenarios

We use the approximate power spectrum (eg. Efstathiou, Bond, & White 1992)

$$|\delta_k|^2 = \frac{Bk^n}{\{1 + [ak + (bk)^{3/2} + (ck)^2]^\nu\}^{2/\nu}} \tag{21}$$

where $a = (6.4/\Gamma)h^{-1}$ Mpc, $b = (3.0/\Gamma)h^{-1}$ Mpc, $c = (1.7/\Gamma)h^{-1}$ Mpc and $\nu = 1.13$. The fitting formula is valid if the baryon density is much smaller than the cold dark matter density. The power spectrum determines the rms fluctuations $\Delta$ on scale $r_0$ through equation (10). We always normalize the power spectrum to the fluctuations $\sigma_8$ on $r_0 = 8h^{-1}$ Mpc, so $\Delta(r_0) = \sigma_8 \hat{\Delta}(r_0)$ and $\hat{\Delta}(8h^{-1}\text{Mpc}) = 1$. The calculation depends on the Hubble constant $h$, the power spectrum normalization $\sigma_8$, the shape of the power spectrum ($n$ and $\Gamma$), and the cosmological model. Before considering individual models, we consider the sensitivity of the lens calculation to the various parameters.

The lens calculations are insensitive to the value of the Hubble constant. The number density of galaxies as a function of velocity dispersion (eqn. 9) depends on $h^3$, but it is multiplied by the $h^{-3}$ dependence of the comoving volume element in the probability calculation (eqn. 6). The Hubble constant dependence of the $a$, $b$, and $c$ coefficients of the power spectrum, excluding the parameter $\Gamma$, gives the integral a dependence on $1/(hr_0) \propto 1/\sigma$, again canceling the dependence on the Hubble constant. When we normalize the spectrum by $\sigma_8$ we remove the Hubble constant dependence of the normalization constant $B$. Thus the expected number of lenses depends on the Hubble constant only through changes in the shape parameter $\Gamma$, and this dependence is weak for $0.5 < h < 1.0$. The COBE estimates for $\sigma_8$ are sensitive to the Hubble constant (eg. Efstathiou, Bond, & White 1992, Bond 1994), so we generally show the lens estimates for $h = 0.5$ compared to the COBE estimates for $h = 0.5$ and $h = 1.0$.

When we study cluster lenses with separations larger than $5''\!.0$, we are examining objects with velocity dispersions between 300 km s$^{-1}$ and the largest objects that can collapse and virialize before the present epoch. The velocity dispersion $\sigma$ for a perturbation of scale $r_0$ is

$$\sigma(r_0) = 950 \left[\frac{r_0}{8h^{-1}\text{Mpc}}\right] (1+z)^{1/2} \text{ km s}^{-1} \tag{22}$$

in an $\Omega_0 = 1$ cosmology (eqn. 16). Thus the scale used to normalize the power spectrum ($r_0 = 8h^{-1}$ Mpc) is comparable to the scale producing the lenses, and the number of large separation lenses is controlled by $dn/d\sigma \propto \sigma_8^{-3} \exp(-\delta_c^2/2\sigma_8^2)$ at the $8h^{-1}$ Mpc scale. The expected number of lenses depends exponentially on the normalization when $\sigma_8 < \delta_c$, and the number varies as a power law when $\sigma_8 > \delta_c$. Since the critical overdensity is $\sigma_c = 1.69(1+z)$ for $\Omega_0 = 1$, the number of lenses will vary exponentally with $\sigma_8$. The smaller separation lenses produced by galaxies correspond to smaller, more nonlinear perturbations where $dn/d\sigma$ is in the power-law regime.

For a fixed value of $\sigma_8$ the number of lenses depends weakly on the shape of the power spectrum. The slope of the power spectrum controls the the separation distribution of the lenses. Power spectra that increase more rapidly for smaller wavenumbers will produce lens separation distributions that decline more steeply with increasing image separation than flatter power spectra. The maximum likelihood method can be made sensitive to the separation distribution by using the probability that a lens has a given separation (eqn. 7) in the lensed term of the likelihood (eqn. 8) instead of the probability that it is a lens (eqn. 6). Given the small number of lenses in the current sample this will not strongly constrain the shape of the power spectrum, but we include the separation probability distribution in one of two standard statistical models to get a feeling for how strongly it will constrain the slope of the power spectrum.

### 5.1 Standard CDM

In standard cold dark matter (CDM) models, the primordial scale-invariant spectrum is $|\delta_k|^2 = Bk$ ($n = 1$), the cosmology is flat with $\Omega_0 = 1$, and the parameter $\Gamma = h$ (eg. Efstathiou et al. 1992). The only free parameters are the normalization constant $B$ which we parametrize by the variance in the density $\sigma_8$ on the scale of $r_0 = 8h^{-1}$ Mpc and the Hubble constant $h$. The measurements of the microwave background fluctuations by the COBE DMR experiment (Smoot et al. 1992, Wright et al. 1994, Górski et al. 1994) require $Q_{rms-PS} = (19.9 \pm 1.6)\mu K$ if $n = 1$ which corresponds to $\sigma_8 \simeq (2.8 \pm 0.2)h$.

Wambsganss et al. (1994) numerically calculated the optical depth to lensing for a standard CDM model with $\sigma_8 = 1.05$ and $h = 0.5$. They found that the optical depth for lenses larger than $10\farcs0$ ($5\farcs0$) with flux ratios smaller than $\Delta m = 1.5$ magnitudes are 0.0007 (0.0008), 0.0014 (0.0019), and 0.0020 (0.0027) for source redshifts of $z_s = 1$, 2, and 3 respectively. The PS calculation with the same parameters and selection function yields cross sections of 0.0004 (0.0006), 0.0012 (0.0017), and 0.0019 (0.0027). The differences in the optical depths are 43% (31%), 21% (11%), and 10% (0%). Narayan & White (1988) used an approximation for $\Delta(r_0)$ that accounts for most of the differences noted by Cen et al. (1994) in their comparison to the PS model. Cen et al. (1994) predict 46 lenses larger than $4\farcs0$ in the older Hewett & Burbidge (1989) catalog for $\sigma_8 = 1$ including magnification bias, comparable to the 27 predicted here even though the samples, quasar number counts, and selection model are different.

The separation distribution of lenses (see Figure 9) in the PS simulations declines more slowly than the numerical simulations (for $\sigma_8 = 1.05$), but the distributions are similar for $\Delta\theta \lesssim 40''$. Detailed comparisons are difficult because Wambsganss et al. (1994) use the largest detected image separation, and this procedure will transfer power from large separations to smaller separations because of the five image systems with only two detectable images (see §6). The separation scale of the deviations roughly corresponds to perturbations with comoving scales significantly larger than the size of the high resolution numerical simulations ($5h^{-1}$ Mpc). The integral redshift distribution of the lenses in the PS model declines more rapidly at low redshifts than in the numerical simulations. This can be explained by adding a core radius comparable to the grid resolution of the numerical simulations ($10h^{-1}$ kpc) in our lens potential model. We consider the effects of ellipticity and core radii in more detail in §6. Nonetheless, the two calculations are everywhere in agreement by a factor of two or better when there is an appreciable cross section for lensing and on scales smaller than $40\rlap{.}''0$. Moreover, the exponential dependence of the number of lenses on $\sigma_8$ means that large uncertainties in the optical depth, the selection function, and the number of lenses lead to much smaller uncertainties in the value of $\sigma_8$. For the standard CDM model, changing the expected number of lenses by a factor of two near $\sigma_8 = 1.4$ changes the estimate of $\sigma_8$ by 0.4, and changing it by a factor of two near $\sigma_8 = 0.5$ changes the estimate of $\sigma_8$ by less than 0.1. We discuss the sources of systematic errors in §6.

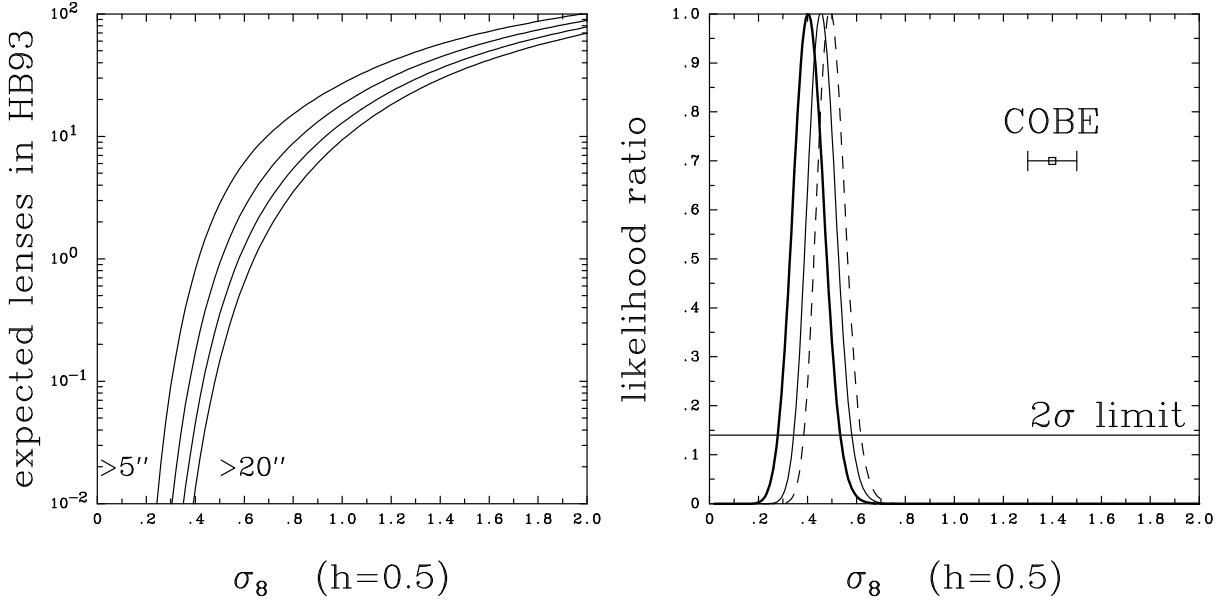

Fig. 3.—Lenses in standard CDM. The left panel shows the expected numbers of lenses larger than $5\farcs0$, $10\farcs0$, $15\farcs0$, and $20\farcs0$ as a function of $\sigma_8$ for $h = 0.5$. Standard CDM normalized to COBE requires $\sigma_8 = (2.8 \pm 0.2)h$. The right panel shows the likelihood ratio as a function of $\sigma_8$ for lenses larger than $5\farcs0$. The heavy solid curve assumes Q0957+561 is the only lens, the light solid curve assumes Q0957+561 and Q2345+007 are lenses, and the dashed curve assumes Q0957+561, Q2345+007, and Q1343+264 are lenses. The horizontal line shows the $2\sigma$ limit on the likelihood ratio, and the point labeled COBE shows the best COBE normalization for $\sigma_8$ ($h = 0.5$) and its $1\sigma$ error bar.

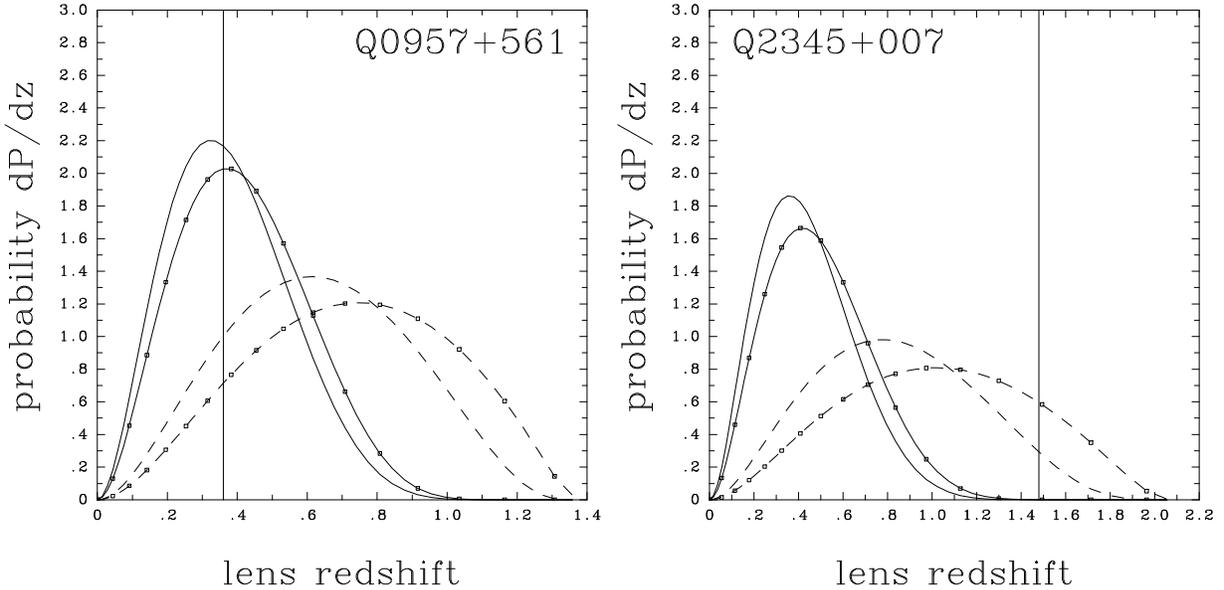

Fig. 4.—Lens redshifts in standard CDM. The two panels show the probability distributions for the lens redshift given the observed image separations in Q0957+561 and Q2345+007. The dashed lines use the COBE normalized values of $\sigma_8$ and the solid lines have $\sigma_8 = 0.5$. The lines with no points use a Hubble constant of $h = 0.5$, and the lines with points use a Hubble constant of $h = 1.0$. The vertical lines mark the known lens redshift in Q0957+561 and a suggested redshift for Q2345+007 (Fischer et al. 1994).

Figure 3 shows the expected number of lenses as a function of $\sigma_8$ in standard CDM. The number of lenses rises exponentially with the bias $1/\sigma_8$, and 60 lenses larger than $5\rlap{.}''0$ are expected at the COBE normalization of $\sigma_8 = 1.4 \pm 0.1$ for $h = 0.5$. This is coincidentally similar to the cross section calculations of Wambsganss et al. (1994) because an order of magnitude error in the estimated cross section is balanced by an order of magnitude error from neglecting the magnification bias. Figure 3 also shows the likelihood ratio as a function of $\sigma_8$ if we include the one real lens (Q0957+561), the real lens and the best candidate (Q0957+561 and Q2345+007), or the real lens, the best candidate, and the rejected candidate (Q0957+561, Q2345+007, and Q1343+264). The $2\sigma$ range for $\sigma_8$ is $0.27 < \sigma_8 < 0.63$, setting the lower limit using one lens and the upper limit with all three. The maximum likelihood values are $\sigma_8 = 0.40$, 0.46, and 0.50 for one, two, and three lenses. These limits are little changed by reducing the separation limit to $3\rlap{.}''0$ or changing the Hubble constant to $h = 1$.

The lens results rule out standard CDM normalized by COBE. The best fit value for $\sigma_8$ is consistent with other estimates based on number density of clusters (Peebles et al. 1989, Frenk et al. 1990, Bahcall & Cen 1992, 1993) or the correlation function of galaxies on large scales (Maddox et al. 1990, Picard 1991, Vogeley et al. 1992, Loveday et al. 1992). It is, however, an independent test controlled by the properties of the cluster distribution at moderate redshifts rather than today. Figure 4 shows the lens redshift probability distributions for Q0957+561 and Q2345+007. The lens redshift in Q0957+561 is known to be $z_l = 0.36$, and it lies at the peak probability of the probability distribution normalized to produce the observed number of lenses ($\sigma_8 \simeq 0.5$). The COBE normalized distributions extend to much higher redshifts because with larger values of $\sigma_8$ the clusters form earlier. In Q2345+007, Fischer et al. (1994) suggest that the lens redshift is $z_l = 1.48$ because there are many metal absorption features at that redshift. In COBE normalized CDM $z_l = 1.48$ is a plausible cluster redshift, but in the models that predict the observed numbers of lenses it is extremely unlikely.

## 5.2 Tilted CDM

The standard scale-free inflationary model assumes that the primordial power spectrum is $|\delta_k|^2 \propto k$. One set of models designed to solve the problems standard CDM has in simultaneously fitting the COBE scale and smaller scale structures, "tilt" the CDM spectrum by making the primordial spectrum $|\delta_k|^2 \propto k^n$ with $n \neq 1$ (eg. Vittorio et al. 1988, Cen et al. 1992). Current fits to the COBE data (Wright et al. 1994, Górski et al. 1994) find that $n = 1.2 \pm 0.3$, consistent with $n = 1$. The fitted value for the $Q_{rms-ps}$ parameter varies with the value of $n$, and Górski et al. (1994) give $Q_{rms-ps} = 18.7(1 + 0.05n) \exp(0.73(1 - n))$ $\mu$K for the best fit value as a function of $n$. Combined with the fitting formula for estimating $\sigma_8$ from $Q_{rms-ps}$ by Bond (1994) we find

$$\sigma_8 = (2.8 \pm 0.2)(\Gamma - 0.03)(1 + 0.05n) \exp\left[-1.9(1 - n)\right]. \qquad (23)$$

This normalization neglects any tensor-mode contributions from gravitational waves. Figure 5 shows contours of the likelihood of fitting the lens data as a function of $\sigma_8$ and the exponent of the tilt $n$. The primordial spectrum must be flatter than $n = 1$ for the lens estimate of $\sigma_8$ to agree with COBE, simply because a flatter spectrum gives less small scale power on the $\sigma_8$ scale when the larger scale is fixed by COBE.

For $h = 0.5$ the lens and COBE limits agree if $0.3 \lesssim n \lesssim 0.7$ ($2\sigma$), and if $h = 1.0$ they agree if $0.0 \lesssim n \lesssim 0.3$ ($2\sigma$). The lens calculations were done with $h = 0.5$, but they are insensitive to the difference between $h = 0.5$ and $h = 1.0$, so we can effectively make the comparison using the $h = 0.5$ calculation. The tilt exponents at the upper end of the allowed range for $h = 0.5$ are marginally consistent (at $2\sigma$) with the COBE limits on the value of $n$, although adding a tensor-mode/gravitational wave contribution to the COBE signal would improve the agreement. The left panel of Figure 5 only includes lenses with separations larger than $5''\!.0$. Since the likelihood has no information other than the expected number of lenses, it cannot differentiate between different values for the exponent $n$ without outside information on $\sigma_8$. Tilting the spectrum rearranges the distribution of image separations, and flat spectra ($n < 1$) produce relatively more large separation lenses than steeper spectra ($n = 1$). If we normalize $\sigma_8$ so that we find three lenses larger than $5''\!.0$, then for $n = 0$, $n = 1$, and $n = 2$ we expect 0.6, 0.3, and 0.1 lenses larger than $10''\!.0$ (see Figure 9).

We can add sensitivity to the value of $n$ by using the probability that the lenses have their observed separations, and the right panel of Figure 5 uses the modified likelihood for the lenses larger than $3''\!.0$. By comparing the two estimates we draw three conclusions. First, the estimated range for $\sigma_8$ does not markedly change when we alter the angular selection function. Second, if there is only one observed lens we have no leverage to determine the value of $n$ because even for the whole range of $0 < n < 2$ the separation of Q0957+561 never becomes unlikely. Third, if we add even a small number of additional lenses, the likelihood contours begin to close. If Q0957+561, Q2345+007, and Q1635+267 are all lenses, then values of $n$ smaller than $n = 0.4$ are ruled out at the $1\sigma$ level. When $n \sim 0$ the expected separation distribution is too flat for all three lenses to have separations between $3''\!.0$ and $7''\!.0$ (see Figure 9).

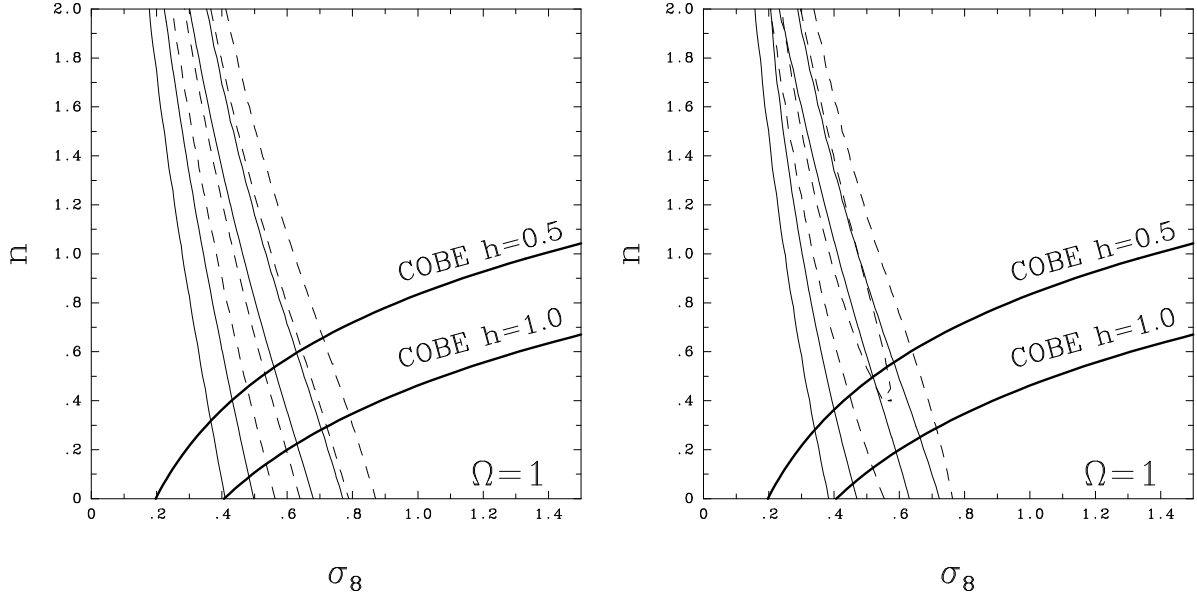

FIG. 5.—Lenses in tilted CDM. The left panel uses the likelihood of finding lenses with separations larger than 5″.0. The solid lines are likelihood contours for finding one lens (Q0957+561) and the dashed lines are likelihood contours for finding three lenses (Q0957+561, Q2345+007, and Q1343+264) as a function of the exponent $n$ and the $\sigma_8$ normalization. The right panel uses the likelihood of finding lenses with separations larger than 3″.0 and the likelihood that the observed lenses have their measured separations. The solid lines are likelihood contours for finding one lens (Q0957+561) and the dashed lines are likelihood contours for finding three lenses (Q0957+561, Q2345+007, and Q1635+267) as a function of the exponent $n$ and the $\sigma_8$ normalization. The contours are the $1\sigma$ and $2\sigma$ confidence intervals of the likelihood ratio for one parameter. The best fit models are bounded by the contours. The heavy solid lines show the COBE normalized estimates for $\sigma_8$ when $h = 0.5$ and $h = 1.0$.

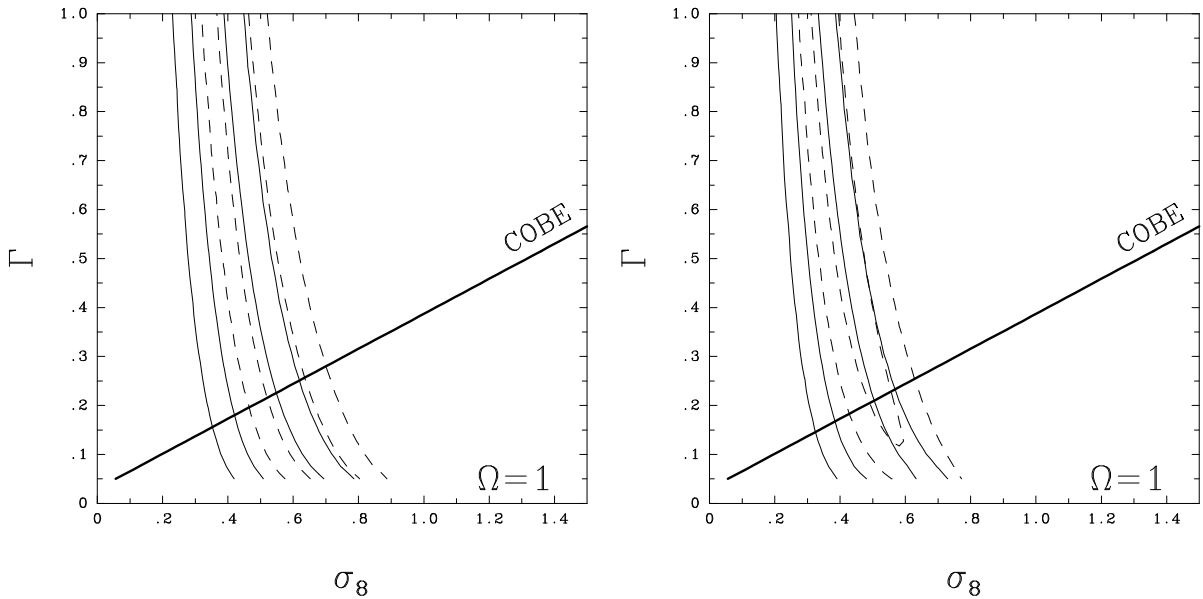

FIG. 6.—Lenses in fitted CDM. The left panel uses the likelihood of finding lenses with separations larger than 5″.0, and the right panel uses the likelihood of finding lenses with separations larger than 3″.0 and the likelihood the observed lenses have their measured separations. The heavy solid lines show the COBE normalized estimates for $\sigma_8$, and the cases and contour levels are the same as in Figure 5.

### 5.3 The Hubble Constant And Varying $\Gamma$

In the standard power spectrum with $\Omega_0 = 1$, $\Gamma = h$ depends only on the Hubble constant. This is the only place where the calculation depends on the value of the Hubble constant. We extend the calculations over a wider range for $\Gamma$ than is plausible for the Hubble constant, and in this broader treatment $\Gamma$ is just a fitting parameter in the power spectrum. Figure 6 shows the dependence of the likelihood on $\Gamma$ and $\sigma_8$ for both $\Delta\theta > 5\rlap{.}''0$ with the total lens probabilities and $\Delta\theta > 3\rlap{.}''0$ with the lens separation probability. When $\Gamma$ is increased from 0.5 to 1.0 the best fit values of $\sigma_8$ shift downwards by 0.1. Since the shift in $\sigma_8$ over the range $0.5 < h < 1.0$ for the Hubble constant is significantly smaller than the error bars, we can regard lens models with $h = 0.5$ as being equivalent to the same model with $h = 1.0$. There is no plausible value of the Hubble constant that can bring the lens and COBE estimates into agreement, but if we regard $\Gamma$ simply as a fitting parameter, they are consistent when $0.15 \lesssim \Gamma \lesssim 0.30$ ($2\sigma$) and $0.3 \lesssim \sigma_8 \lesssim 0.7$. These ranges are similar to the ranges found from the correlation function on scales near $10h^{-1}$ Mpc (Maddox et al. 1990, Picard 1991, Vogeley et al. 1992, Loveday et al. 1992) and from observed numbers of large clusters (Bahcall & Cen 1992, 1993). The lens estimate for $\sigma_8$ as a function of $\Gamma$ is a measure of the rms perturbations in the mass distribution, so the agreement with estimates of $\sigma_8$ from luminous objects implies that there is little or no bias on cluster scales. With the inclusion of the lens separations, the likelihood contours begin to pinch off for small values of $\Gamma$ (see Figure 9).

### 5.4 Open and Flat Low $\Omega_0$ Universes

Reducing the matter density has three different effects on lens statistics. The first effect is that lower density universes have larger comoving volumes to a given redshift, so for a constant comoving density of objects we expect more lenses. An empty universe has twice as many lenses (Turner, Ostriker, & Gott 1984) and a flat universe with $\lambda_0 = 1$ has ten times as many lenses (Turner 1990) as a flat universe with $\Omega_0 = 1$. This allows us to use the incidence of galaxy scale lenses to determine the cosmological model (eg. Kochanek 1993b, Maoz & Rix 1993). The second effect is that perturbations stop growing at low redshifts (at $z \sim \Omega_0^{-1}$ for open universes) so clusters must collapse earlier in low matter density universes if they are to be seen today. Richstone et al. (1992), Lacey & Cole (1993), Bartelmann et al. (1993) use this to argue that high values of $\Omega_0$ are needed to explain the observed substructure and rapid evolution of clusters at low redshifts. Both of these effects work to increase the expected number of lenses. The third effect is that a perturbation on scale $r_0$ contains less mass in a low $\Omega_0$ universe, and when it collapses it tends to have a larger virial radius. The mass of the perturbation scales with $\Omega_0 r_0^3$, and the virial radius is proportional to the maximum expansion radius.

For a fixed value of $\sigma_8$, the third effect is the dominant one, and lowering $\Omega_0$ reduces the number of observable lenses. Because the velocity dispersion of the collapsed perturbation for a fixed value of $r_0$ is smaller, lenses with a fixed image separation are produced by perturbations on larger and larger comoving scales as $\Omega_0$ is reduced. At fixed $\sigma_8$, this means that the rms fluctuation amplitude of the perturbations producing a fixed image separation decreases. This makes changes in $\Omega_0$ for a fixed $\sigma_8$ similar to changes in $\sigma_8$ for a fixed $\Omega_0$, because of the way it changes the power spectrum normalization on the scales probed by lenses with a fixed separation. Since the number of lenses is exponentially sensitive to the normalization, the third effect dominates over the other two. This is very different from galaxy scale lenses, where the geometric effects of changing the cosmological model are more important than evolution (Mao 1991, Mao & Kochanek 1994, Rix et al. 1994).

Figure 7 shows the likelihood contours for open universes compared to the COBE normalizations determined for $\Omega_0 < 1$ CDM models by Kamionkowski & Spergel (1994) renormalized to fit the Górski et al. (1994) rms quadrupole estimates. The approximate $2\sigma$ limit is $0.15 \lesssim \Omega_0 h \lesssim 0.3$, although this overestimates the upper limit for $h = 0.5$ by about 0.1 in $\Omega_0$. Note that when we add the separation probabilities and three lenses, the likelihood contours have pinched off at both low $\Omega_0$ giving a nominal one standard deviation limit of $\Omega_0 \gtrsim 0.25$. Obviously this is a weak conclusion, but it shows how the separation distribution of a larger sample of large separation lenses can constrain the slope of the power spectrum. Figure 8 shows the likelihood for flat universes with a cosmological constant. The limits are consistent with the COBE $Q_{rms-ps}$ if $0.45 \lesssim \lambda_0 \lesssim 0.85$ ($2\sigma$) for $h = 0.5$ and $0.85 \lesssim \lambda_0 \lesssim 1.0$ ($2\sigma$) for $h = 1.0$. We again see signs that the observed separation distribution may be too steep for low values of $\Omega_0$. Cen et al. (1994) also find that models with $\lambda_0 = 0.3$ have nearly an order of magnitude fewer lenses than models with $\lambda_0 = 0.0$ when $\sigma_8 = 1$. Recall, however, that gravitational lensing by galaxies strongly rules out large values of the cosmological constant (the two standard deviation upper limit is $\lambda_0 < 0.6$, Kochanek 1994), and that the observed, low value of the microwave background quadrupole compared to $Q_{rms-ps}$ supports this conclusion (Sugiyama & Silk 1994).

## 6   Systematic Errors

Before concluding we should discuss some of the systematic errors in this calculation. We can divide the problems into shortcomings with the PS method as a means of estimating the number of halos, shortcomings in converting a collapsed perturbation on scale $r_0$ into a virialized object with velocity dispersion $\sigma$, and shortcomings in modeling the lenses as circular, singular isothermal spheres. To illustrate the effects of these systematic errors, we examined the standard CDM model to see how the estimate of $\sigma_8$ depends on changes in each of these parts of the model.

The critical part of the PS model is $\delta_c(z)$, the critical overdensity collapsing at redshift $z$. While errors in the details of the lens model and the selection function cause only logarithmic changes in estimates of $\sigma_8$, changes in $\delta_c$ lead to proportionate changes in $\sigma_8$ because the expected number of lenses depends roughly on $\exp(-\delta_c^2/2\sigma_8^2)$. If we vary $\delta_c$ between 0.5 and 1.5 of its standard value, then the estimated value of $\sigma_8$ changes by $\pm 0.2$ for finding one lens and by $\pm 0.25$ for finding three lenses. Reducing $\delta_c$ lowers the perturbation amplitude needed to produce the observed number of lenses. This scaling is roughly as expected, since a 50% variation in $\delta_c$ produces a 50% variation in the best fit value of $\sigma_8$. If we multiply $\delta_c$ by 1.5 and assume a sample with three lenses, then the the $2\sigma$ upper limit on $\sigma_8$ rises to 0.92.

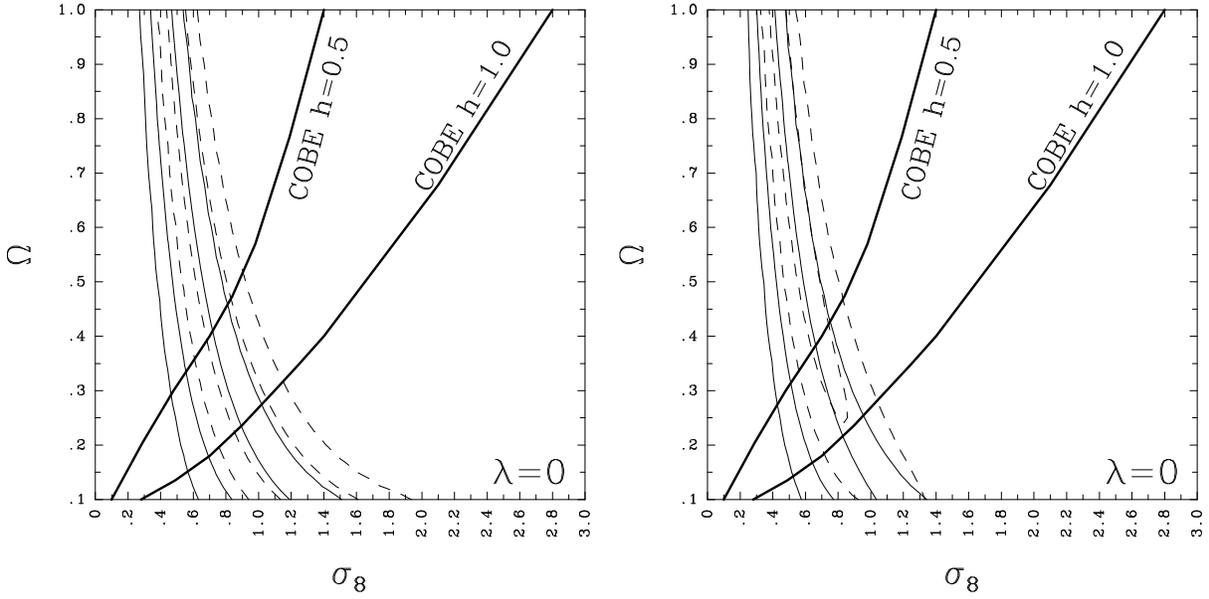

FIG. 7.–Lenses in low $\Omega_0$ CDM. The left panel uses the likelihood of finding lenses with separations larger than $5\rlap.{''}0$, and the right panel uses the likelihood of finding lenses with separations larger than $3\rlap.{''}0$ and the likelihood the observed lenses have their measured separations. The heavy solid lines show the COBE normalized estimates for $\sigma_8$ from Kamionkowski & Spergel (1994), and the cases and contour levels are the same as in Figure 5.

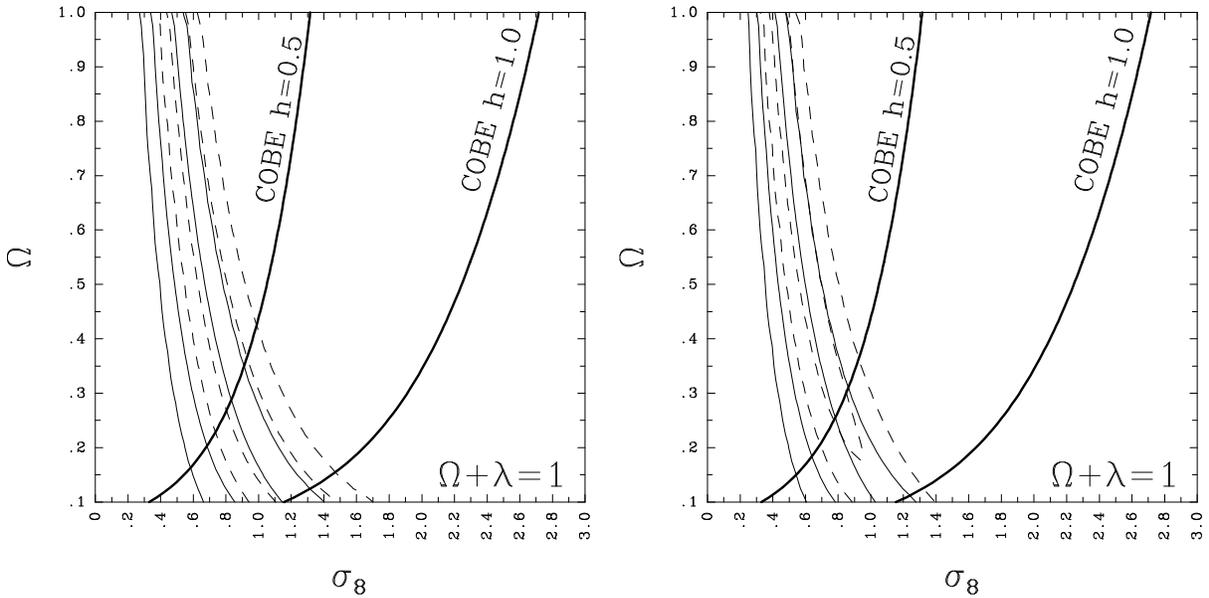

FIG. 8.–Lenses in $\Omega_0 + \lambda_0 = 1$ CDM. The left panel uses the likelihood of finding lenses with separations larger than $5\rlap.{''}0$, and the right panel uses the likelihood of finding lenses with separations larger than $3\rlap.{''}0$ and the likelihood the observed lenses have their measured separations. The heavy solid lines show the COBE normalized estimates for $\sigma_8$, and the cases and contour levels are the same as in Figure 5.

The virial theorem estimates for converting the perturbation scale $r_0$ into a velocity dispersion are also crude. The estimate of $\sigma_8$ is sensitive to the estimated ratio of the final virialized radius because by changing the relation between $r_0$ and $\sigma$ we shift the comoving scale that produces lenses of a given separation. Lowering the final virial radius means that a given velocity dispersion can be produced by smaller perturbations, reducing the estimate of $\sigma_8$. We used the estimate (for $\lambda_0 = 0$) that the final virialized radius was one half the maximum expansion radius $R_f = R_{max}/2$, and the velocity dispersion estimate is proportional to the square root of the ratio $\sigma \propto (R_{max}/R_f)^{1/2}$. For example, the velocity dispersion estimate rises by 10% if we treat the collapsed perturbations as Jaffe (1983) models and normalize the velocity dispersion by conserving energy instead of mass. If we halve the virial radius $R_f = R_{max}/4$ then the velocity dispersion associated with a perturbation is 40% larger, and the best fit values of $\sigma_8$ are reduced by 0.13 for one lens and by 0.19 for two lenses, while if we double the virial radius $R_f = R_{max}$ then the velocity dispersion is 40% lower, and the best fit values of $\sigma_8$ are increased by 0.23 for one lens and by 0.32 for three lenses. If we take three lenses and increase the virial radius by a factor of two, the $2\sigma$ upper limit on $\sigma_8$ is 1.14. The value of $\sigma_8$ is roughly scaling as $(2R_f/R_{max})$.

If we add ellipticity to the lens model we do not change the average cross section of the lenses but we radically alter the range of image morphologies because the lenses produce four image as well as two image systems. This can alter the number of observable lenses through changes in the magnification bias and the detectability of the new image configurations. We generated Monte Carlo samples of $10^4$ lenses produced by singular isothermal spheres in an external shear field with dimensionless ellipticity $\gamma = 0.1$ (see Kochanek 1991) at 15, 16, 17, 18, and 19 B magnitudes. The average magnification bias before applying the selection function is nearly identical to the circular model. We then applied the selection function to the Monte Carlo catalogs, using the average estimate for the limiting magnitude ($\langle \Delta m \rangle$) and found that the expected number of lenses was within 20% of the estimates for the circular lens. This has a negligible effect on estimates of $\sigma_8$. The introduction of elliptical lenses does radically alter the expected morphologies because the four image systems are almost always detected – the incompleteness comes entirely from reducing the numbers of two image systems. Thus at 16, 17, 18, and 19 B mags the expected fraction of four image lenses is 87%, 45%, 33%, and 30% of the total, even though they consist of only 63%, 41%, 13%, and 4% of the lenses at those magnitudes. The preponderance of the four image systems is not too overwhelming to be inconsistent with finding only two images systems in the existing sample. Moreover, particularly for the brighter systems, roughly 85% of the four image systems are only detectable as image pairs in a quasar survey, with the remaining two images lying beyond the magnitude limit. Surveys tend to find the quads as merging pairs on a critical line because they are substantially magnified and have similar fluxes. The introduction of ellipticity has no interesting effect on estimates of $\sigma_8$.

The last systematic problem we consider is the addition of a core radius to the lens model. We know from the models of giant arcs (see the review by Soucail & Mellier 1994) that the core radii of the arc producing clusters are compact enough to produce well separated tangential and radial critical lines for comparatively low redshift sources ($z_s \lesssim 1$), and we know from studies of the galaxy scale lenses that the core radii of galaxies are very compact (eg. Wallington & Narayan 1993, Kassiola & Kovner 1993), but we have no direct information on the core radii of groups and clusters with intermediate masses. We emphasize comparisons to limits from lensing because they are the most direct constraints on the core radii used in a lensing calculation. Numerical simulations of structure formation uniformly lead to collapsed objects with core radii set by the smoothing length of the simulation, consistent with all objects having nearly singular dark matter halos (eg. Dubinski & Carlberg 1991). When we consider adding core radii to the models we must consistently include the modifications of the magnification bias, because correcting the expected number of lenses using only the changes in the lens cross section produced by core radii leads to serious quantitative errors. If the lenses have the density distribution $\rho \propto (r^2 + s^2)^{-1}$ then the lenses are subcritical if

$$\frac{2s}{b} = 0.05 \frac{s_{10}}{\sigma_3^2} \frac{D^A_{OS} r_H}{D^A_{OL} D^A_{LS}} > 1 \qquad (24)$$

where the $D^A$ are angular diameter distances, $s = 10 s_{10} h^{-1}$ kpc, and $b = 4\pi(\sigma/c)^2 D^A_{LS}/D^A_{OS}$ is the critical radius of the singular model (Hinshaw & Krauss 1987). For $\Omega_0 = 1$ and a source at redshift $z_s = 2$, the core radius must be smaller than $s < 28 h^{-1} \sigma_3^2$ kpc for the lens to produce multiple images. The $10 h^{-1}$ kpc comoving grid scale in the Wambsganss et al. (1994) simulations is dangerously close to this limit. If we assume that the core radii of clusters scale with $s \propto \sigma^2$ so that the ratio $s/b$ is independent of velocity dispersion and all clusters have an equal capacity for lensing, then for a source at $z_s = 2$ we find that the expected number of lenses is reduced by a factor of two for a core radius of $s = 15 h^{-1} \sigma_3^2$ kpc, it is reduced by a factor of ten for a core radius of $s = 19 h^{-1} \sigma_3^2$ kpc, and it is zero for $s > 28 h^{-1} \sigma_3^2$ kpc, including the effects of magnification bias. For intermediate core radii of order $7 h^{-1} \sigma_3^2$ kpc the expected number of lenses is approximately 50% higher than for a singular lens because the average magnification is slightly higher. For lower redshift sources the limits on the core radius will be somewhat smaller, and for higher redshift sources the limits will be somewhat higher.

Although a large core radius can be used to escape the lensing constraints, such an assumption leads to some contradictions. If the core radius is tuned to reduce the number of lenses by an order of magnitude, then we expect all observed lenses to be marginal lenses. Marginal lenses generally have visible central or "odd images," and their absence can be used to set limits on the core radius (Wallington & Narayan 1993, Kassiola & Kovner 1993). In the observed sample of large separation quasar lenses we do not see any central images, suggesting that the core radii of clusters must be near singular. An alternate, and logically dangerous, argument is to reverse the direction of argument and note that when we normalize the cosmogonic model based on other observations (COBE, cluster number counts, and correlation functions), we only produce the observed numbers of lenses if the clusters are assumed to be nearly singular.

## 7 CONCLUSIONS

The key to making a quantitative comparison between observations of large separation gravitational lenses and cosmological predictions is a good model for the selection effects that determine whether the lenses are found by quasar surveys. Based on the simple picture that quasar surveys must have optical magnitude limits, we built a selection function model that explains the magnitude distribution of wide separation pairs in the HB93 catalog. One important implication of the model is that quasar surveys are a very inefficient method of finding wide separation quasar pairs.

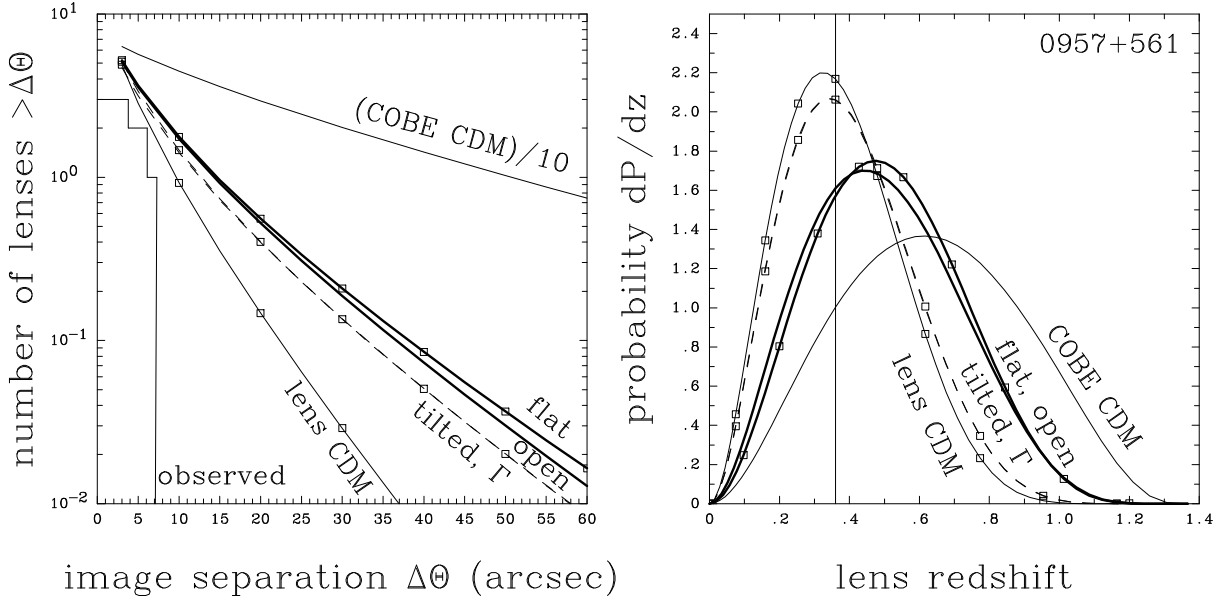

FIG. 9.—The integral separation distributions (left panel) and the differential lens redshift probability distribution for 0957+561 (right panel) for various models with $h = 0.5$. The solid line (COBE CDM) shows the distributions for COBE normalized CDM ($\sigma_8 = 1.4$) and the solid line with points (lens CDM) shows the distributions for CDM normalized to produce 3 lenses larger than $5\rlap{.}''0$ ($\sigma_8 = 0.5$). In the left panel, the number of lenses produced by the COBE normalized CDM model is divided by a factor of 10 to fit it on the same scale as the other models. The figure also shows a tilted model ($n = 0.6$, $\sigma_8 = 0.6$, dashed lines), a $\Gamma$ model ($\Gamma = 0.25$, $\sigma_8 = 0.6$, dashed lines with points), an open model ($\Omega_0 = 0.45$, $\sigma_8 = 0.8$, heavy solid lines), and a flat model with a cosmological constant ($\Omega_0 = 0.35$, $\sigma_8 = 0.9$, heavy solid lines with points) that produce approximately 3 lenses larger than $5\rlap{.}''0$ and are consistent with COBE. The tilted model and the $\Gamma$ model give nearly identical results. The histogram in the left panel shows the integral separation distribution of the known lens Q0957+561 and the two best candidates Q2345+007 and Q1635+667. The vertical line in the right panel marks the known redshift of the lens in 0957+561.

We use the Press-Schechter (1974) approximation to compute the expected number of lenses in different cosmogonic scenarios, an approach pioneered by Narayan & White (1988). When we compare the PS results to the Wambsganss et al. (1994) numerical simulations we find that the overall agreement is good. For example, the difference in the expected number of lenses with separations larger than $10\rlap{.}''0$ for a source at reshift $z_s = 2$ is only 20%. Such small differences have negligible effects on estimates of $\sigma_8$, the rms fluctuation over a $8h^{-1}$ Mpc top-hat window function, because the number of lenses depends exponentially on $\sigma_8$. Some differences between the PS models and the Wambsganss et al. (1994) simulations are clearly traceable to differences in the models. The lens redshift distributions differ because we use a singular isothermal sphere for the lens model, while the simulations have mass distributions with finite core radii ($\sim 10h^{-1}$ kpc) set by the resolution of the calculation. When we compute the redshift distribution with a $10\text{-}20h^{-1}$ kpc core radius for the lenses we reproduce the numerical simulations. For large separation lenses the PS results predict many more wide separation lenses, and the deviations between the two results begin with separations of order $40\rlap{.}''0$. This is approximately the image separation scale corresponding to perturbations comparable to the outer scale of the numerical simulations ($5h^{-1}$ Mpc). Other differences between the models are not easily assigned to either approach.

Systematic errors in the PS approach, either in estimating the critical overdensity for collapse, $\delta_c(z)$, or the virial radius of the collapsed perturbation do not strongly modify the conclusions. The estimates of $\sigma_8$ vary linearly with rescaling either $\delta_c$ or $\sigma/r_0$, so rescaling either of these relations by 50% rescales the estimate of $\sigma_8$ by 50% at $\sigma_8 \simeq 0.5$. Changing the expected number of lenses by a factor of two either by altering the selection function model or the structure of the lenses changes the estimate of $\sigma_8$ by 0.1 for $\sigma_8 \simeq 0.5$. Adding ellipticity to the lens model leads to a small change in the expected number of lenses, and many (30% to 50%) of the lenses detected by quasar surveys will consist of two images of a four image lens. Since the actual number of large separation lenses is small and uncertain (at least one, possibly three) most of theses systematic errors change the estimated value of $\sigma_8$ by amounts comparable to the intrinsic statistical uncertainties.

The addition of a large core radius to the density distributions of clusters and groups can strongly affect the conclusions. A core radius larger than $15h^{-1}\sigma_3^2$ kpc reduces the expected number of lenses by a factor of two, and a core radius larger than $19h^{-1}\sigma_3^2$ reduces the expected number of lenses by a factor of ten. Simulations of dark matter halos (eg. Dubinski & Carlberg 1991), galaxy scale lenses (Wallington & Narayan 1993, Kassiola & Kovner 1993), and the models of giant arcs (see Soucail & Mellier 1994) all suggest that potentials tend to have compact core radii. Moreover there is a serious fine tuning problem in giving groups and clusters core radii large enough to allow consistency with COBE normalized CDM, while producing only a few lenses and making the observed lenses have undetectable odd images in the lens cores. Neither the PS simulations nor the numerical simulations are capable of addressing this issue in any detail, but a larger sample of lenses will strongly constrain the core radii.

We determined the values of $\sigma_8$ required to produce the observed number of wide separation lenses in standard CDM, tilted CDM, low $\Omega_0$ cosmologies, and flat cosmologies with a cosmological constant. Figure 9 summarizes the distributions of lens separations and the probability distribution for the lens redshift for standard CDM and various models normalized to fit both COBE and the lens data. The general results agree with other estimates of $\sigma_8$ in the various models. For standard CDM models we find $\sigma_8 \simeq 0.45 \pm 0.2$ ($2\sigma$) for $h = 0.5$ similar to estimates from correlation functions (Maddox et al. 1990, Picard 1991, Vogeley et al. 1992, Loveday et al. 1992) and cluster abundances (Peebles et al. 1989, Frenk et al. 1990, Bahcall & Cen 1992, 1993), and in strong disagreement with the normalization $\sigma_8 = (2.8 \pm 0.2)h$ needed to fit the COBE observations (Smoot et al. 1992, Wright et al. 1994, Górski et al. 1994). As previously noted by Narayan & White (1988), Cen et al. (1994), and Wambsganss et al. (1994), COBE normalized CDM overpredicts the number of lenses by more than an order of magnitude. It also predicts a much flatter distribution of image separations than is observed, and higher average lens redshifts (see Figure 9). Tilted models with a primordial spectrum $\propto k^n$ with $0.3 \lesssim n \lesssim 0.7$ ($h = 0.5$) are consistent with the COBE estimates for $\sigma_8$ at those values of the exponent $n$. The limits on the exponent $n$ from the COBE data are $n = 1.2 \pm 0.3$ (Wright et al. 1994, Górski et al. 1994), so the overall consistency is poor. Changes in the Hubble constant have negligible effects on the lens models, with the best fit value for $\sigma_8$ decreasing by 0.1 when we increase $h$ from 0.5 to 1.0. Treated purely as a fitting function in the power spectrum, effective Hubble constants (the $\Gamma$ parameter in the power spectrum) of $0.15 \lesssim \Gamma \lesssim 0.30$ are consistent with the COBE data and the observed number of lenses. All the variant models that produce the observed number of lenses and are consistent with COBE have steeper distributions of lens separations and lower average lens redshifts than standard CDM (see Figure 9). Since lensing determines $\sigma_8$ of the mass distribution, the fact that it agrees with estimates of $\sigma_8$ from luminous objects suggests that there is little or no bias on cluster scales. We did not examine cold + hot models, although they should show better agreement because they reduce the numbers of groups and clusters (eg. Nolthenius et al. 1994).

Models with low matter densities whether flat with a cosmological constant or open models are consistent with COBE and the lens observations if $0.3 \lesssim \Omega_0 \lesssim 0.5$ for $h = 0.5$. Unlike galaxy scale lenses where evolution appears to be unimportant (Mao 1991, Mao & Kochanek 1991, Rix et al. 1994), evolution matters far more than the extra volume of the low matter density universes. The same number of lenses is produced by higher values of $\sigma_8$ largely because the mass of a collapsed perturbation for a fixed comoving scale $r_0$ decreases rapidly when $\Omega_0$ is reduced. The high cosmological constant models are strongly ruled out by gravitational lensing (Kochanek 1993b, Maoz & Rix 1993, Kochanek 1994) with a current two standard deviation limit of $\lambda_0 \lesssim 0.6$. The low value of the COBE quadrupole compared to the higher $\ell$ fluctuations sets a weaker limit that $\lambda_0 \lesssim 0.8$ (Sugiyama & Silk 1994). Moreover, if the cosmological constant is invoked to solve the age problem produced by a high value of the Hubble constant, the value of $\lambda_0$ is forced to be larger than 0.85 exacerbating the conflict with the other limits.

In the models normalized to produce the observed number of lenses, the average completeness of the lens sample is 20%. In the best fitting models, 90% of the lenses are smaller than $20\rlap{.}''0$. The quasar surveys miss the lenses because of their limited average dynamic range, and the existing lens surveys are poorly designed for finding wide separation lenses. Lens surveys examine regions $5''$ to $10''$ in radius around each quasar for lensed images using one or two color photometry followed by spectroscopy. The survey region is limited because the background of galactic stars is already several times the number of lenses within these small regions. A wide separation lens survey is equivalent to a new, deep quasar survey centered on a known quasar, with the same selection problems and contamination problems of quasar surveys. If the goal is restricted to finding lensed images of a known quasar, the optimal solution for surveying large areas around the quasar is to use intermediate-width filters ($200\text{Å}$ wide) bracketing the strongest available emission line of the quasar. Since the equivalent width of a strong quasar emission line is a large fraction of the filter bandwidth, there should be very few objects other than quasars at the same redshift capable of matching the color of the quasar.

Acknowledgements: The author thanks M. Bartelmann, D. Eisenstein, A. Loeb, R. Narayan, and D. Spergel for discussions about this paper.REFERENCES

Arp, H., Sulentic, J.W., & Di Tullio, G., 1979, ApJ, 229, 489

Arp, H., 1983, ApJ, 271, 479

Arp, H., & Surdej, J., 1982, A&A, 109, 101

Bahcall, N., & Cen, R., 1992, ApJ, 398, L81

Bahcall, N., & Cen, R., 1993, ApJ, 407, L49

Bartelmann, M., Ehlers, J., & Schneider, P., 1993, A&A, 280, 351

Bond, J.R., Cole, S., Efstathiou, G., & Kaiser, N., 1991, ApJ, 379, 440

Bond, J.R., 1994, in *Relativistic Cosmology*, M. Sasaki, ed., (Academic Press)

Boyle, B.J., Shanks, T., & Peterson, B.A., 1988, MNRAS, 235, 935

Boyle, B.J., Fong, R., Shanks, T., & Peterson, B.A., 1990, MNRAS, 243, 1

Burke, B., Lehàr, J., & Conner, S.R., 1992, in *Gravitational Lenses*, R. Kayser, T. Schramm, & L. Nieser, eds., (Springer: Berlin) 237

Carroll, S.M., Press, W.H., & Turner, E.L., 1992, ARA&A, 30, 499

Cen, R., Gnedin, N.Y., Kofman, L.A., & Ostriker, J.P., 1992, ApJ, 399, L11

Cen, R., & Ostriker, J., 1992, ApJ, 393, 22

Cen, R., Gott, J.R., Ostriker, J.P., & Turner, E.L., 1994, ApJ, 423, 1

Crampton, D., Cowley, A.P., Hickson, P., Kindl, E., Wagner, R.M., Tyson, J.A., & Gullixson, C., 1988, ApJ, 330, 184

Djorgovski, S., & Spinrad, H., 1984, ApJ, 282, L1

Djorgovski, S., Perley, R., Meylan, G., & McCarthy, P., 1987, ApJ, 321, L17

Djorgovski, S., & Meylan, G., 1989, in *Gravitational Lenses*, J. Moran, J. Hewitt, & K.-Y. Lo, eds., (Berlin: Springer) 173

Dubinski, J., & Carlberg, R.G., 1991, ApJ, 378, 496

Efstathiou, G., Bond, J.R., & White, S.D.M., 1992, MNRAS, 258, 1P

Fischer, P., Tyson, J.A., Bernstein, G.M., & Guhathakurta, P., 1994, ApJ, 431, L71

Frenk, C.S., White, S.D.M., Efstathiou, G., & Davis, M., 1990, ApJ, 351, 10

Fukugita, M., & Turner, E.L., 1991, MNRAS, 253, 99

Górksi, K.M., Hinshaw, G., Banday, A.J., Bennett, C.L., Wright, E.L., Kogut, A., Smoot, G.F., & Lubin, P., 1994, ApJ, 430, L89

Gott, J.R., & Gunn, J.E., 1974, ApJ, 190, L105

Hartwick, F.D.A., & Schade, D., 1990, ARA&A, 28, 437

Hewett, P.C., Webster, R.L., Harding, M.E., Jedrzejewski, R.I., Foltz, C.B., Chaffee, F.H., Irwin, M.J., & Le Fèvre, O., 1989, ApJ, 346, L61

Hewitt, A., & Burbidge, G., 1993, ApJS, 87, 225

Hewitt, A., & Burbidge, G., 1989, ApJS, 69, 1

Hinshaw, G., & Krauss, L.M., 1987, ApJ, 320, 468

Kamionkowski, M., & Spergel, D.N., 1994, ApJ, 432, 7

Kassiola, A., & Kovner, I., 1993, ApJ, 417, 450

Kochanek, C.S., 1991, ApJ, 379, 517

Kochanek, C.S., 1993a, MNRAS, 261, 453

Kochanek, C.S., 1993b, ApJ, 419, 12

Kochanek, C.S., 1994, in *Critique of the Sources of Dark Matter in the Universe*, D. Cline, ed., (World Scientific)

Jaffe, W., 1983, MNRAS, 202, 995

Lacey, C., & Cole, S., 1993, MNRAS, 262, 627

Lahav, O., Lilje, P.B., Primack, J.R., & Rees, M.J., 1991, MNRAS, 251, 128

Lawrence, C.R., Schneider, D.P., Schmidt, M., Bennett, C.L., Hewitt, J.N., Burke, B.F., Turner, E.L., & Gunn, J.E., 1984, Science, 223, 46

Loveday, J., Efstathiou, G., Peterson, G.A., & Maddox, S.J., 1992, ApJ, 400, L43

Lupton, R. L., 1993, *Statistics in Theory and Practice*, (Princeton University Press: Princeton)

Mao, S. 1991, ApJ, 380, 9

Mao, S., & Kochanek, C.S., 1994, MNRAS, 268, 569

Maoz, D., Bahcall, J.N., Schneider, D.P., Bahcall, N.A., Djorgovski, S., Doxsey, R., Gould, A., Kirhakos, S., Meylan, G., & Yanny, B., 1993, ApJ, 409, 28

Maoz, D., & Rix, H.-W., 1993, ApJ, 416, 425

Maddox, S.J., Efstathiou, G., Sutherland, W.J., & Loveday, J., 1990, MNRAS, 242, 43P

Meylan, G., & Djorgovski, S., 1989, ApJ, 338, L1

Narayan, R., & White, S.D.M., 1988, MNRAS, 231, 97P

Nolthenius, R., Klypin, A., & Primack, J.R., 1994, ApJ, 422, L45

Peebles, P.J.E., 1980, *The Large-Scale Structure of the Universe* (Princeton University Press: Princeton)

Peebles, P.J.E., Daly, R.A., & Juszkiewicz, R., 1989, ApJ, 347, 563

Picard, A., 1991, AJ, 102, 445

Press, W.H., & Schechter, P., 1974, ApJ, 187, 425

Richstone, D., Loeb, A., & Turner, E.L., 1992, ApJ, 393, 477

Rix, H.-W., Maoz, D., Turner, E.L., & Fukugita, M., 1994, *IASSNS-AST 94/16 preprint*

Smoot, G.F., et al. 1992, ApJ, 396, L1

Soucail, G., & Mellier, Y., 1994, in *Gravitational Lenses in the Universe*, J. Surdej, D. Fraipont-Caro, E. Gosset, S. Refsdal, & M. Remy, eds., (University of Liege) 595

Sugiyama, N., & Silk, J., 1994, PhysRevLett, 73, 509

Surdej, J., & Soucail, G., 1994, in *Gravitational Lenses in the Universe*, J. Surdej, D. Fraipont-Caro, E. Gosset, S. Refsdal, & M. Remy, eds., (University of Liege) 205

Surdej, J., Claeskens, J., Crampton, D., Filippenko, A.V., Hutsemékers, D., Magain, P., Pirenne, B., Vanderriest, C., & Yee, H.K.C., 1993, AJ, 105, 2064

Turner, E.L., 1980, ApJ, 242, L135

Turner, E.L., Ostriker, J.P., & Gott, J.R., 1984, ApJ, 284, 1

Turner, E.L., 1990, ApJ, 365, L43

Vittorio, N., Matarrese, S., & Lucchin, F., 1988, ApJ, 328, 69

Vogeley, M.S., Park, C., Geller, M.J., & Huchra, J., 1992, ApJ, 391, L5

Wallington, S., & Narayan, R., 1993, ApJ, 403, 517

Walsh, D., Carswell, R.F., Weymann, R.J., 1979, Nature, 279, 381

Wambsganss, J., Cen, R., Ostriker, J.P., & Turner, E.L., 1994, *Princeton preprint*

Webster, R.L., Hewett, P.C., & Irwin, M.J., 1988, AJ, 95, 19

Weedman, D.W., Weymann, R.J., Green, R.F., & Heckman, T.M., 1982, ApJ, 255, L5

Weymann, R.J., Latham, D., Angel, J.R.P., Green, R.F., Liebert, J.W., Turnshek, D.A., Turnshek, D.E., & Tyson, J.A., 1980, Nature, 285, 641

Wright, E.L, Smoot, G.F., Kogut, A., Hinshaw, G., Tenorio, L., Lineweaver, C., Bennett, C.L, & Lubin, P.M., 1994, ApJ, 420, 1